\documentclass[twocolumn, tighten, dvipsnames, floatfix]{aastex63}

\usepackage[T1]{fontenc}
\usepackage[]{xcolor}
\usepackage{booktabs}
\usepackage{physics}
\usepackage{wasysym}

\usepackage{longtable}
\usepackage{threeparttablex} 

\usepackage{array}
\usepackage{amsmath}
\usepackage{changepage}
\usepackage{lineno, ulem} 
\usepackage[flushmargin]{footmisc}
\usepackage{afterpage}
\usepackage{subfigure}

\newcommand{\kms}{\mbox{$\>{\rm km\, s^{-1}}$}}

\def\arcsec{\hbox{$^{\hbox{\rlap{\hbox{\lower4pt\hbox{$\,\prime\prime$}}
          }}}$} \ }

\def\arcmin{\hbox{$^{\hbox{\rlap{\hbox{\lower4pt\hbox{$\;\prime$}}
          }\hbox{$\frown$}}}$}}

\usepackage{lipsum}
\usepackage{float}

\shorttitle{The Monoceros Overdensity}
\shortauthors{Borbolato et al.}

\begin{document}

\title{All-sky Kinematics and Chemistry of Monoceros Stellar Overdensity}

\correspondingauthor{Lais Borbolato}
\email{laisborbolato@gmail.com}

\author[0000-0003-3382-1051]{Lais Borbolato}
\affil{Universidade de S\~ao Paulo, Instituto de Astronomia, Geof\'isica e Ci\^encias Atmosf\'ericas, Departamento de Astronomia, \\ SP 05508-090, S\~ao Paulo, Brasil}

\author[0000-0002-0537-4146]{H\'elio D. Perottoni}
\affil{Universidade de S\~ao Paulo, Instituto de Astronomia, Geof\'isica e Ci\^encias Atmosf\'ericas, Departamento de Astronomia, \\ SP 05508-090, S\~ao Paulo, Brasil}
\affil{Nicolaus Copernicus Astronomical Center, Polish Academy of Sciences, ul. Bartycka 18, 00-716, Warsaw, Poland}

\author[0000-0001-7479-5756]{Silvia Rossi}
\affil{Universidade de S\~ao Paulo, Instituto de Astronomia, Geof\'isica e Ci\^encias Atmosf\'ericas, Departamento de Astronomia, \\ SP 05508-090, S\~ao Paulo, Brasil}

\author[0000-0002-9269-8287]{Guilherme Limberg}
\affil{Universidade de S\~ao Paulo, Instituto de Astronomia, Geof\'isica e Ci\^encias Atmosf\'ericas, Departamento de Astronomia, \\ SP 05508-090, S\~ao Paulo, Brasil}
\affil{Department of Astronomy \& Astrophysics, University of Chicago, 5640 S. Ellis Avenue, Chicago, IL 60637, USA}
\affil{Kavli Institute for Cosmological Physics, University of Chicago, 5640 S. Ellis Avenue, Chicago, IL 60637, USA}

\author[0000-0002-5974-3998]{Angeles P\'erez-Villegas}
\affil{Instituto de Astronom\'ia, Universidad Nacional Aut\'onoma de M\'exico, Apartado Postal 106, C. P. 22800, Ensenada, B. C., M\'exico}

\author[0000-0003-4524-9363]{Friedrich Anders}
\affil{Departament de F\'isica Qu\'antica i Astrof\'isica (FQA), Universitat de Barcelona (UB),  Mart\'i i Franqu\'es, 1, 08028 Barcelona, Spain}
\affil{Institut de Ci\'encies del Cosmos (ICCUB), Universitat de Barcelona (UB), Mart\'i i Franqu\'es, 1, 08028 Barcelona, Spain}
\affil{Institut d'Estudis Espacials de Catalunya (IEEC), Gran Capit\'a, 2-4, 08034 Barcelona, Spain}

\author[0000-0000-0000-0000]{Teresa Antoja}
\affil{Departament de F\'isica Qu\'antica i Astrof\'isica (FQA), Universitat de Barcelona (UB),  Mart\'i i Franqu\'es, 1, 08028 Barcelona, Spain}
\affil{Institut de Ci\'encies del Cosmos (ICCUB), Universitat de Barcelona (UB), Mart\'i i Franqu\'es, 1, 08028 Barcelona, Spain}
\affil{Institut d'Estudis Espacials de Catalunya (IEEC), Gran Capit\'a, 2-4, 08034 Barcelona, Spain}

\author[0000-0000-0000-0000]{Chervin F. P. Laporte}
\affil{Departament de F\'isica Qu\'antica i Astrof\'isica (FQA), Universitat de Barcelona (UB),  Mart\'i i Franqu\'es, 1, 08028 Barcelona, Spain}
\affil{Institut de Ci\'encies del Cosmos (ICCUB), Universitat de Barcelona (UB), Mart\'i i Franqu\'es, 1, 08028 Barcelona, Spain}
\affil{Institut d'Estudis Espacials de Catalunya (IEEC), Gran Capit\'a, 2-4, 08034 Barcelona, Spain}

\author[0000-0002-5274-4955]{Helio J. Rocha-Pinto}
\affiliation{Universidade Federal do Rio de Janeiro, Observat\'orio do Valongo, Lad. Pedro Ant\^onio 43, 20080-090, Rio de Janeiro, Brazil}

\author[0000-0002-7529-1442]{Rafael M. Santucci}
\affil{Universidade Federal de Goi\'as, Instituto de Estudos Socioambientais, Planet\'ario, Goi\^ania, GO 74055-140, Brazil}
\affil{Universidade Federal de Goi\'as, Campus Samambaia, Instituto de F\'isica, Goi\^ania, GO 74001-970, Brazil}


\begin{abstract}
 
We explore the kinematic and chemical properties of Monoceros stellar overdensity by combining data from 2MASS, WISE, APOGEE, and \text{Gaia}. Monoceros is a structure located towards the Galactic anticenter and close to the disk. We identified that its stars have azimuthal velocity in the range of $200 < v_{\phi}\,{\rm(km\,s^{-1})}< 250$. Combining their kinematics and spatial distribution, we designed a new method to select stars from this overdensity. This method allows us to easily identify the structure in both hemispheres and estimate their distances. Our analysis was supported by comparison with simulated data from the entire sky generated by \texttt{Galaxia} code. Furthermore, we characterized, for the first time, the Monoceros overdensity in several chemical-abundance spaces. Our results confirm its similarity to stars found in the thin disk of the Galaxy and suggest an \textit{in situ} formation. Furthermore, we demonstrate that the southern (Mon-S) and northern (Mon-N) regions of Monoceros exhibit indistinguishable chemical compositions.
\end{abstract}




\section{Introduction}
\label{sec:intro}

Several attempts to understand the structure and formation history of the Milky Way (MW) arise from observations and measurements of the kinematics and chemistry from stars that surround us \citep{Eggen1962, Gilmore1983, Majewski1993, Freeman2002, BlandHawthorn2016} as well as sparse tracers of more distant parts of the Galaxy \citep[e.g.,][]{Majewski2003, Hernitschek2017, Iorio2019}. These observations led to the currently most accepted scenario that the MW had a bottom-up formation (\citealt{searle1978,spergel2007, somerville2015}), including old and recent mergers of dwarf galaxies, which formed substructures from the tidal disruption of these satellites \citep{Helmi2020}. Despite these efforts to understand in detail the MW formation history, it remains unclear how and when some structures were formed, such as stellar streams \citep{Majewski2003, belokurov2006, Bonaca2012, Shipp2018, Malhan2018} and overdensities \citep{newberg2002, Rocha_Pinto2003, ibata2003, belokurov2007a, xu2015, Li2016mgiants}.

Overdensities are Galactic regions identified as stellar count excesses when compared with homogeneous regions. The study of these regions makes it possible to identify events that occurred during the evolution of the MW that caused their formation, contributing to assembling the pieces of the puzzle of how the Galaxy evolved. Overdensities can be found both in the Galactic halo and disk. Those in the halo  like Virgo \citep{newberg2002,juric2008}, Hercules-Aquila \citep{belokurov2007a}, Pisces \citep{sesar2007,watkins2009} and Eridanus-Phoenix \citep{Li2016mgiants} have their origin associated with accretion events that occurred in the MW \citep{Simion2019, Balbinot2021, Naidu2021, Perottoni2022}. On the other hand, the structures close to the disk, such as Monoceros \citep{newberg2002},  Triangulum-Andromeda \citep{rocha2004, majewski2004}, Anticenter Stream \citep{grillmair2006}, and several structures identified in \citet{laporte2022}, seem to have a Galactic origin \citep[e.g.,][]{bergemann2018}.

One of the most studied stellar overdensities is Monoceros. Also named Galactic Anticenter Stellar Structure (\citealt{Rocha_Pinto2003}), 
it is a stellar overdensity which can be seen as having two distinct parts: one in the Northern Galactic Hemisphere and the other in the Southern Galactic Hemisphere. Both parts are located in a region of low Galactic latitude in the direction of the Galactic Anticenter (120$^{\circ} < l <$ 240$^{\circ}$; \citealt{morganson2016}).
\citet{newberg2002} discovered this structure using a two-dimensional polar density histogram with data from the Sloan Digital Sky Survey (SDSS; \citealt{york2000}). In their work, three pieces named S223+20$-$19.4, S218+22$-$19.5, and S183+22$-$19.4 were identified in the northern hemisphere as being part of the same structure, located at distance of approximately 11 kpc from the Sun. S200$-$24$-$19.8 in the southern hemisphere was associated with the same structure, at a distance of 8 kpc from the Sun. 
Other studies also identified Monoceros in both Galactic hemispheres at different distances (e.g., \citealt{Rocha_Pinto2003, xu2015, morganson2016}). In spite of the fact that both parts of the structure were discovered approximately at the same time, the Monoceros in the northern hemisphere has been explored with greater details than its counterpart in the south due to a lack of coverage in photometric and spectroscopic surveys.

The study and characterization of the Monoceros led to different origin scenarios, which include \textit{in situ} and \textit{ex situ} hypotheses. \citet[][using data from SDSS]{newberg2002} and \citet[][with data from Two Micron All Sky Survey; 2MASS; \citealt{skrutskie2006}]{Rocha_Pinto2003} explored the possibilities that Monoceros could be either part of a disk population or the result of the tidal disruption of a dwarf galaxy, and support the latter hypothesis (see also \citealt{crane2003}).  
Later, Monoceros was associated with a possible remnant core of a dwarf galaxy in Canis Major \citep{Martin2004} and in Argo (\citealt{Rocha-pinto2006}). However, it was suggested that the star counts in the direction of Canis Major and Monoceros could be reproduced by incorporating warp and flare into the stellar density description of the Galactic disk models  \citep{momany2004, momany2006, moitinho2006, LopezCorredoira2007}, supporting the disk origin scenario. On the same side of this debate, \cite{xu2015} identified that Monoceros, and other stellar overdensities, could be produced by oscillations of the disk midplane, which differ from the hypothesis that the disc warp and flare could solely account for the nature of Monoceros. This hypothesis was supported by \cite{price2015} and \cite{sheffield2018} whose analyses pointed out that the number ratio of RR Lyrae to M-giant stars in the stellar overdensities near the plane (e.g. Monoceros, Triangulum-Andromeda) is similar to the Galactic disk. The intense debate between the \textit{in situ} (e.g.: \citealt{LopezCorredoira2007, hammersley2011, Lopez&Molgo2014, sheffield2018}) and \textit{ex situ} (e.g.: \citealt{Penarrubia2005, Rocha-pinto2006, chou2010, morganson2016}) origin continued during the following decade.

Evidence of a possible \textit{in situ} origin was supported by predictions from numerical simulations that demonstrate that the interaction of a dwarf galaxy, such as Sagittarius, with the Milky Way, could create vertical perturbations and overdensities \citep{purcell2011, gomez2016, laporte2018}, which is, currently, the most well-accepted hypothesis for the origin of Monoceros and other overdensities.

Despite numerous efforts to unravel the scenario for the Monoceros formation, chemical analyses of its stars aimed at confirming these hypotheses have been relatively unexplored \cite{chou2010} analysed the chemical abundance patterns from high-resolution spectroscopy of 21 stars in the Monoceros (only from the Northern Galactic Hemisphere) and concluded, from the analysis of $\alpha$, and $s$-process elements, that Monoceros stars have abundance typical of dwarf galaxies. 

In contrast, a more recent analysis of the chemical abundance in the [Fe/H]-[Mg/Fe] plane of northern hemisphere Monoceros stars indicated that the structure is chemically compatible with the thin disk region of the Galaxy \citep{Laporte2020b}. This result reinforces the \textit{in situ} origin scenario related to the interaction of the MW with the Sagittarius dwarf galaxy (Sgr). According \citet{Laporte2020b} Monoceros (north) also presents a cumulative age distribution, which suggests the formation of the structure from multiple passages of Sgr. Unfortunately, the southern portion of Monoceros is yet to be characterized and, hence, a connection with the northern counterpart remains to be confirmed. A homogeneous analysis of the Monoceros in the north and south hemisphere is the next step to better understand its nature and formation. 

In view of the scenario presented above, we aimed in this work to explore the kinematic and chemical properties of Monoceros in the Gaia era, in both hemispheres, in a way never done before. This information may help us answer questions that are still open, such as: Are northern and southern Monoceros connected populations? How do these populations compare to the halo population? 

This paper is organized as follows. In Section \ref{sec:data}, we describe the data utilized in this work. Section \ref{sec:Mon_selection} details the method developed to select candidate stars from the Monoceros overdensity. After obtaining the star sample of the structure, we present an analysis of chemical-abundance information in Section \ref{sec:abundances}. Finally, summarize our results in Section \ref{sec:conclusion}.


\section{Data}
\label{sec:data}

\subsection{M Giants Sample}
\label{sub:MGiants}

\begin{figure*}[ht!]
        \centering
        \includegraphics[width=2.1\columnwidth]{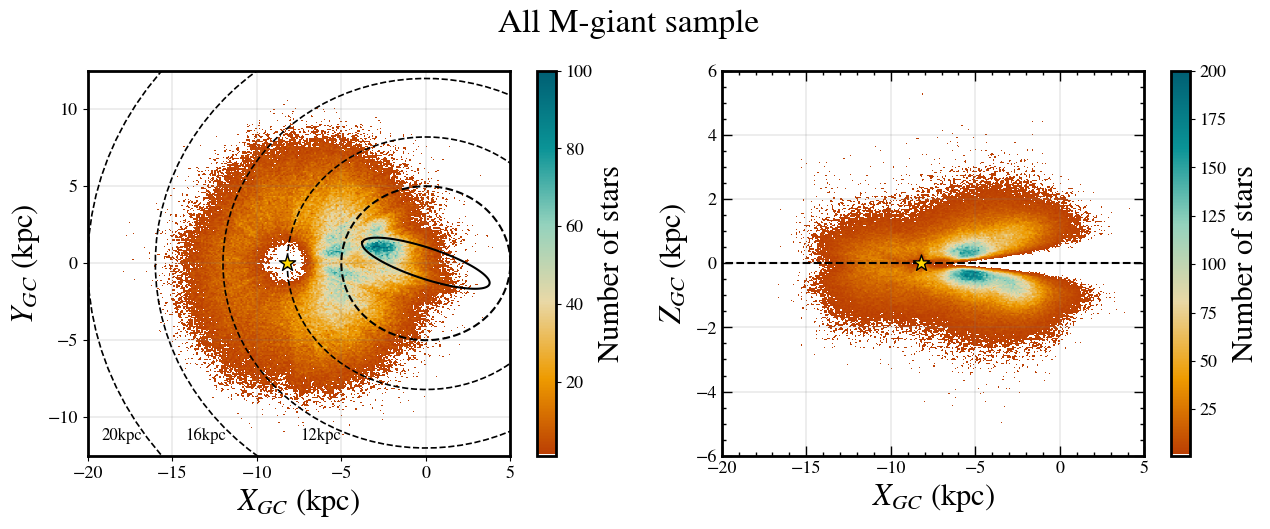}
        \caption{Spatial distribution in the $X_{GC}$--$Y_{GC}$ (left) and $X_{GC}$--$Z_{GC}$ (right) Galactic Cartesian planes of the entire sample of M-giant stars. In the left panel, the dashed lines indicate Galactocentric radii (5\,kpc, 8\,kpc, 12\,kpc, 16\,kpc and 20\,kpc) and the ellipse indicates an approximate position for the bar of the Milky Way \citep{queiroz2021}. In the right panel, the dashed line indicates the Galactic plane ($Z_{GC}$ = 0\,kpc). In all panels, the yellow star indicates the position of the Sun ([$X_{GC}$, $Y_{GC}$] = [$-$8.2, 0.0]\,kpc).}
        \label{XYZ_altura_WISE}
    \end{figure*}

To study the properties of northern and southern Monoceros overdensities, we first need to obtain a large sample of stars from these substructures. 
We have compiled a catalog of M-giant stars to ensure we identify the true members of these substructures beforehand. In addition to being large in numbers, this sample also is a good tracer of structures in the Galaxy \citep{Majewski2003, Rocha_Pinto2003, Sharma2010, price2015}, since M-giant stars are very luminous and can be identified at great distances.

For the purpose of building the M-giant sample, we used the CDS X-match service\footnote{\url{http://cdsxmatch.u-strasbg.fr/}} to cross-match in 1 arcsec radius Point Source Catalog of the 2MASS (\citealt{2MASS,cutri2003}) with All Wide-field Infrared Survery Explore Source Catalog (AllWISE; \citealt{WISE2010,AllWISE2013}). The 2MASS mapped the sky in three photometric near-infrared  bands at 1.2 ($J_{S}$), 1.6 ($H$), and 2.2 ($K_{S}$) microns, whereas the WISE survey obtained data at 3.4 ($W1$), 4.6 ($W2$), 12 ($W3$), and 22 ($W4$) microns. We impose quality cuts to obtain a reliable photometric sample. As we only use the W1 and W2 filters, we select only those sources from AllWISE with no contamination of close objects  ({\tt ccf == 0}), good photometric quality ({\tt qph == A}),  point sources objects ({\tt ext < 2}), not variables ({\tt var < 6}), and not saturated ($W1 >$ 8 mag and $W2 >$ 7 mag). For 2MASS, we select only point sources with good photometric quality ({\tt qfl == A}).

Although the 2MASS photometry has a good efficiency in the selection of M-giant stars, the inclusion of the WISE bands improves the purity of the M-giant selection, as didactically shown in \cite{Koposov2015} and \cite{Li2016mgiants}. Therefore, we adopted the following selection criteria based on \cite{Li2016mgiants} M-giants selection\footnote{We adopted a small change by selecting objects with $(J-K)_{0}$ $>$ 0.8 in order to 
enlarge the M-giant stars catalog to lower metallicities ($[\text{Fe/H}] \approx 0.9$), according to the MIST isochrones \citep{choi2016, dotter2016}.
There is a small intersection with K giants which does not affect the goals of the present work. Furthermore, any possible contamination from nearby dwarf stars is ruled out given the accuracy of the distance estimates adopted in this work.}

\begin{eqnarray}
-0.23<( {\it W1}- {\it W2})_{0}<0.02,\nonumber\\
0.80<(J-K_{S})_{0}<1.3,\label{eq:selection}\\
(J-K)_{0}>1.45\times({\it W1}- {\it W2})_{0}+1.05.\nonumber
\end{eqnarray}

Extinction in $J$, $K_{S}$, $W1$, and $W2$ were calculated as $(A_{J_{S}}, A_{K_{S}}, A_{W1}, A_{W2} = (0.72, 0.306, 0.18, 0.16) \times E_{B-V}$, where $E_{B-V}$ is the reddening in optical bands. The extinction coefficients are from \cite{Yuan2013_ext_coef} and the $E_{B-V}$ is from \cite{Schlegel1998}. In spite of these improvements, an upper limit of $E_{B-V}$ $\leq 0.55$ was imposed on the sample in an effort to prevent differential reddening from affecting the M-giant sample or estimating the reddening inaccurately. As a result of removing high-extinction regions, in combination with the quality and selection cuts in the photometric data from 2MASS and AllWISE, the M-giant sample contains 1,808,525 stars. It is important to point out that this sample does not provide us with chemical abundances from spectroscopic data, as it is formed from photometric catalogs.

Figure \ref{XYZ_altura_WISE} shows the spatial distribution of M-giant sample in the $X_{GC}$--$Y_{GC}$ (left) and $X_{GC}$--$Z_{GC}$ (right) Cartesian-coordinate projections. The noticeable absence of stars around the Sun is primarily due to the removal of saturated stars from the All-WISE data, leading to this artificial feature.
    
\subsection{APOGEE Sample}
\label{sub:apogeesample}

To explore the chemical abundance patterns of the stellar overdensities in detail, we selected stars from the Apache Point Observatory Galactic Evolution Experiment (APOGEE; \citealt{apogee2017}) data release 17 (DR17; \citealt{APOGEEdr17}). Through two identical high-resolution ($R \sim$ 22,000) spectrographs \citep{Wilson2019} installed on 2.5m telescopes in the northern and southern hemispheres \citep{Bowen1973,Gunn2006}, APOGEE has observed the outer disk above and below the Galactic plane in the near infrared (1.51$\mu$m--1.69$\mu$m; \textit{H}-band). The observed targets cover part of the Monoceros North and South stellar overdensities with a good signal-to-noise ratio ($S/N \sim 100$), which allows precise estimates of chemical abundances, radial velocities, and stellar parameters \citep{GarciaPerez2016, jonsson2020}.

We utilized the APOGEE DR17 data to obtain stars with high-quality abundance measurements. To ensure the reliability of the spectroscopic solutions, we applied clean cuts, excluding sources that exhibited any issues in the spectra, the spectral fitting process, and the estimated parameters (\texttt{ASPCAPFLAG} == 0 and \texttt{STARFLAG} == 0; see \citealt{jonsson2020}). Additionally, we ensured accurate estimates of [Fe/H], [Mg/Fe], [Al/Fe], [Mn/Fe], [Ni/Fe], [C/Fe], [N/Fe], and [O/Fe] by selecting sources with no flagged issues (i.e., flagged == 0). Also, we implemented cuts to select only giant stars ($4000 < T{\rm eff} (K) < 6000$, $\log g < 3$). The APOGEE sample contains 149,297 stars.

\subsection{Gaia and StarHorse}

In order to study the phase-space properties of Monoceros, we combined the M-giant and APOGEE samples with Gaia DR3 \citep{GaiaDR32022arXiv} data using a search radius of 3.0$''$ 
to obtain positions, absolute proper motions, and their uncertainties. To ensure high-quality astrometry, we applied selection criteria to the Gaia data.  First, we implemented a selection criteria based on the recommended range of renormalized unit weight errors ($\texttt{RUWE} \leq 1.4$; \citealt{Lindegren2020a}). Additionally, stars with low fidelity ($\texttt{fidelity\_v2} < 0.5$; \citealt{Rybizki2022fidelity}), that could potentially have spurious astrometric solutions, were removed. 
It is worth noting that only stars with available parallax measurements were considered to obtain reliable distance results (see below).

Despite the parallax measurements provided by Gaia, accurate distances for members of stellar overdensities at the outer disk remain challenging to obtain due to their large distances from the Sun \citep{Bailer_Jones2018, Bailer_Jones2021}. To overcome this limitation, we made use of the Bayesian isochrone-fitting code \texttt{StarHorse} \citep{Santiago2016starhorse, Queiroz2018} to obtain reliable distances. Specifically, for our sample of M-giant stars, we obtained photo-astrometric heliocentric distances \citep{Anders2022}. For our APOGEE sample, we used spectro-photo-astrometric distances \citep{Queiroz2020,Queiroz2023}, which takes advantage of spectroscopic information during the isochrone-fitting procedure. These precise distance estimates provided by \texttt{StarHorse} enabled us to confidently identify members of the Monoceros overdensity for a chemo-kinematic analysis of its structure, as demonstrated in its successful application to studying other distant Galactic structures such as stellar streams \citep{Limberg2023sgr}, overdensities \citep{Perottoni2022, Abuchaim2023}, and the bar/bulge \citep{queiroz2021}.

The nominal values for the heliocentric distances of the stars were obtained by adopting the medians of the posterior distributions from \texttt{StarHorse}. The $16$th and $84$th percentiles of the distributions were considered as uncertainties. To have a sufficiently large sample with accurate distance estimates, the sample was limited to stars with fractional Gaussian uncertainties in their nominal distance values below $20\%$. 
This results in a final \textit{M-giant sample} and an \textit{APOGEE sample} with Gaia DR3 of 350,916 and 100,029 stars, respectively.  Our samples have mean fractional distance uncertainties between $12\%$ and $7\%$, indicating their suitability for our analysis.

Finally, we employed Gaia DR3's positions and proper motions on the sky, \texttt{StarHorse} heliocentric distances, and available radial velocities to compute 6D phase-space vector in a Galactocentric Cartesian frame. For the \textit{M-giant sample}, radial velocities from Gaia DR3 were used \citep{GaiaDR3_RVs_2023}. For the \textit{APOGEE sample}, this survey's values were adopted. The assumed position of the Sun with respect to the Galactic Center is $(X,Y,Z)_{GC} = (-8.2, 0.0, 0.0)$ kpc \citep[][see also, e.g., \citealt{GRAVITY2019} and \citealt{Reid2019masers}]{BlandHawthorn2016}. The circular velocity at this position is 232.8\kms \citep{mcmillan2017} and the peculiar motion of the Sun with respect to the circular orbit at its location is $(U,V,W)_\odot = (11.10, 12.24, 7.25)$\kms \citep{schon2010}. 

\subsection{Galactic content with Galaxia}

We made use of the \texttt{Galaxia}\footnote{\url{https://galaxia.sourceforge.net/}} code \citep{sharma2011} with its default parameters to generate a mock catalog of the entire Galaxy. This catalog was created for the purpose of comparing its predicted contents from Galaxia model with the observed data in the Monoceros region. The simulated catalog covers the entire sky up to a magnitude limit of $J < 11.8$, matching the faintest magnitude limit of the observational data. This value is greater than peak of the magnitude distribution observed in the \textit{M-giant sample}, which is J $\sim$ 11. For this reason, the sample of simulated stars is more numerous than the observed stars. Additionally, the simulation accounted for the warp and flare of the Galactic disk. \texttt{Galaxia} provided outputs for magnitudes, positions, kinematical properties, and more, for all stars in the simulation. To reproduce the \textit{M-giant sample}, we selected stars with $0.80 < (J-K_{S})_{0} < 1.3$, applying the same color criteria used for selecting M-giant stars in the observed data, but only for the the 2MASS filters.

\begin{figure*}[ht!]
        \centering
        \includegraphics[width=2.1\columnwidth]{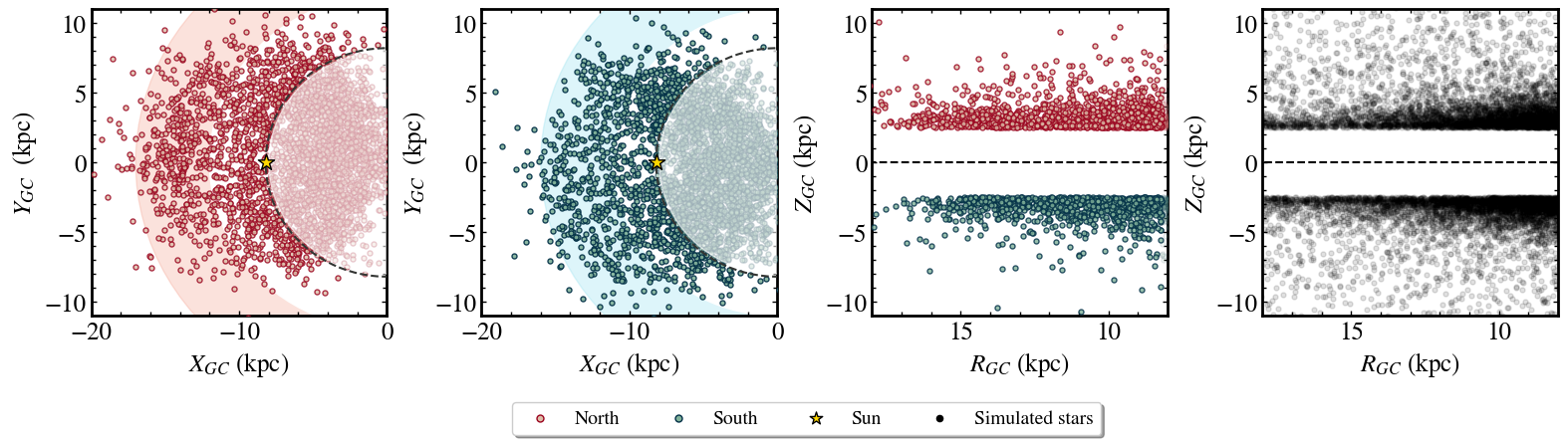}
        \caption{Spatial distribution of M-giant stars after the sample selection in $Z_{GC}$. The first and second panels present the spatial distribution of data in space $X_{GC}$--$Y_{GC}$ for the northern and southern hemispheres, respectively. The shaded area indicates the range of Galactocentric radius associated with Monoceros. The interval is $12 < R_{GC} \text{(kpc)} < 17$ for Mon-N and $11 < R_{GC} \text{(kpc)} < 16$ for Mon-S \citep{newberg2002, Rocha_Pinto2003, morganson2016}. The dashed line indicates the Sun's Galactocentric radius $R_{GC} = 8.2$\,kpc. The third panel indicates the spatial distribution of M-giant stars in the $R_{GC}$--$Z_{GC}$ plane. For comparison, the last panel shows the projection of stars generated by  \texttt{Galaxia} on the $R_{GC}$--$Z_{GC}$ plane. In these last two panels, the dashed line indicates the Galactic plane ($Z_{GC} = 0$\,kpc). Northern hemisphere stars are above 2.5\,kpc in height relative to the Galactic plane and those in the southern hemisphere are below 2.5kpc. The yellow star indicates the position of the Sun ([$X_{GC}$, $Y_{GC}$] = [$-$8.2, 0.0]\,kpc).}
        \label{XYZ_altura}
    \end{figure*}
    

\section{Monoceros Sample Selection}
\label{sec:Mon_selection}


\vspace{0.3cm}

In this section, we describe the Monoceros sample selection. To find structure across both Galactic hemispheres, we use the \textit{M-giant sample} described in Section \ref{sub:MGiants}.

The Monoceros overdensity is a stellar substructure located in both the north and south Galactic hemispheres, likely at different distances. Despite some indications of a possible kinematic connection between the two substructures \citep{Li2017}, studies on the southern component have been scarce, and its association with the northern counterpart through chemical composition is yet to be addressed in the literature. To shed light on these gaps in knowledge and better understand this substructure, we have distinguished Monoceros as two substructures, Monoceros-South (Mon-S) and Monoceros-North (Mon-N).

The Mon-S and Mon-N are situated in regions of low Galactic latitudes, close to the region predominantly occupied by disk stars, making it challenging to identify and select its true members. To overcome this challenge, we obtained a slice in $Z_{GC}$ in the \textit{M-giant sample}, which considerably decreased the number of disk stars in our sample. Therefore, we limited the \textit{M-giant sample} by selecting stars in the northern hemisphere with $Z_{GC} > 2.5$\,kpc (\textit{North M-giant sample}) and stars in the southern hemisphere with $Z_{GC} < -2.5$\,kpc (\textit{South M-giant sample}). This approach is similar to that used by \citet[for instance see their Figure 16]{morganson2016} to map Monoceros. Additionally, we did not impose an upper limit on $Z_{GC}$ values in order to avoid restricting the Galactic latitude of the sample.  Applying this criterion leaves a final \textit{North M-giant sample} and \textit{South M-giant sample} of 3,451 and 3,394 stars, respectively, for our analysis.

Figure \ref{XYZ_altura} shows the spatial distribution of the \textit{M-giant sample}, and of the simulated stars by \texttt{Galaxia}, with the implemented cuts. The first and second panels shows the distribution in the $X_{GC}$--$Y_{GC}$ plane of the \textit{North M-giant sample} and the \textit{South M-giant sample}, respectively. The stars located in the region within the $R_{GC}$ of the Sun are not objects of study in this work, therefore, they are shown with transparency in the figures. In the third panel, it is possible to visualize the distribution of the heights from the Galactic plane of the stars of both hemispheres in the $X_{GC}$--$Z_{GC}$ plane. Finally, the last panel shows the distribution in the $X_{GC}$--$Z_{GC}$ plane of the simulated stars with the same height cuts.

\begin{figure*}[ht!]
        \centering
        \includegraphics[width=2.1\columnwidth]{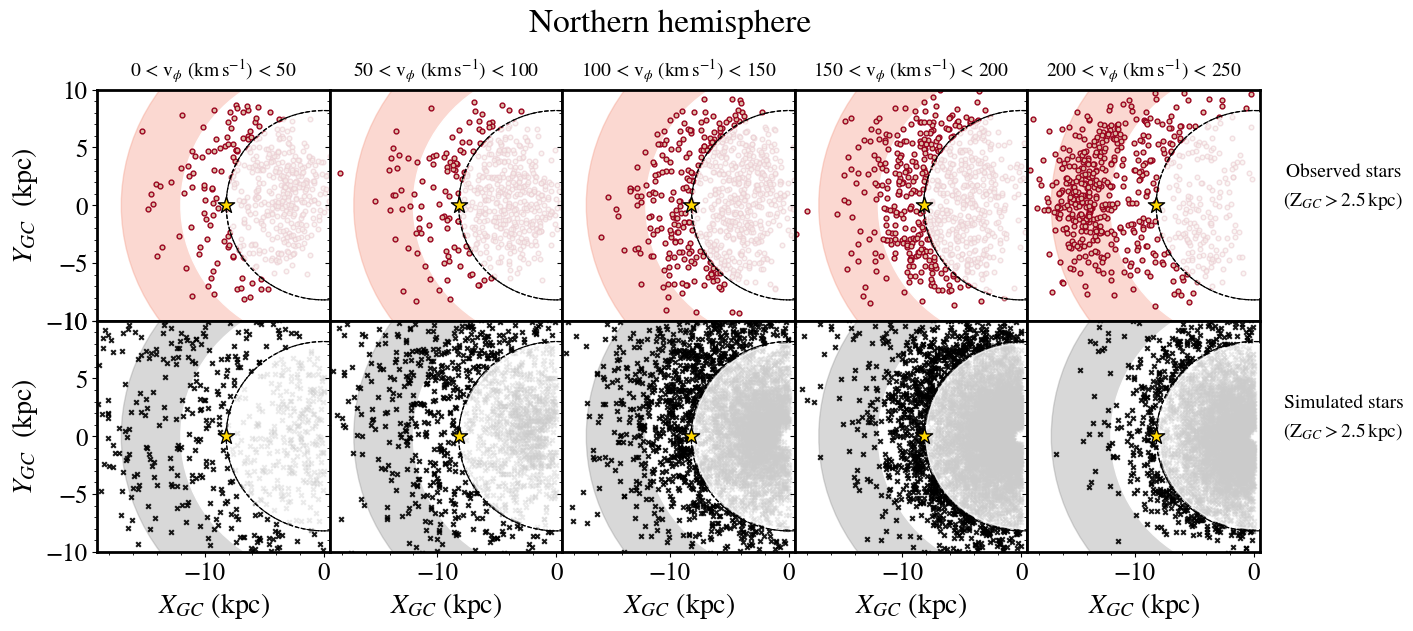}
        \includegraphics[width=2.1\columnwidth]{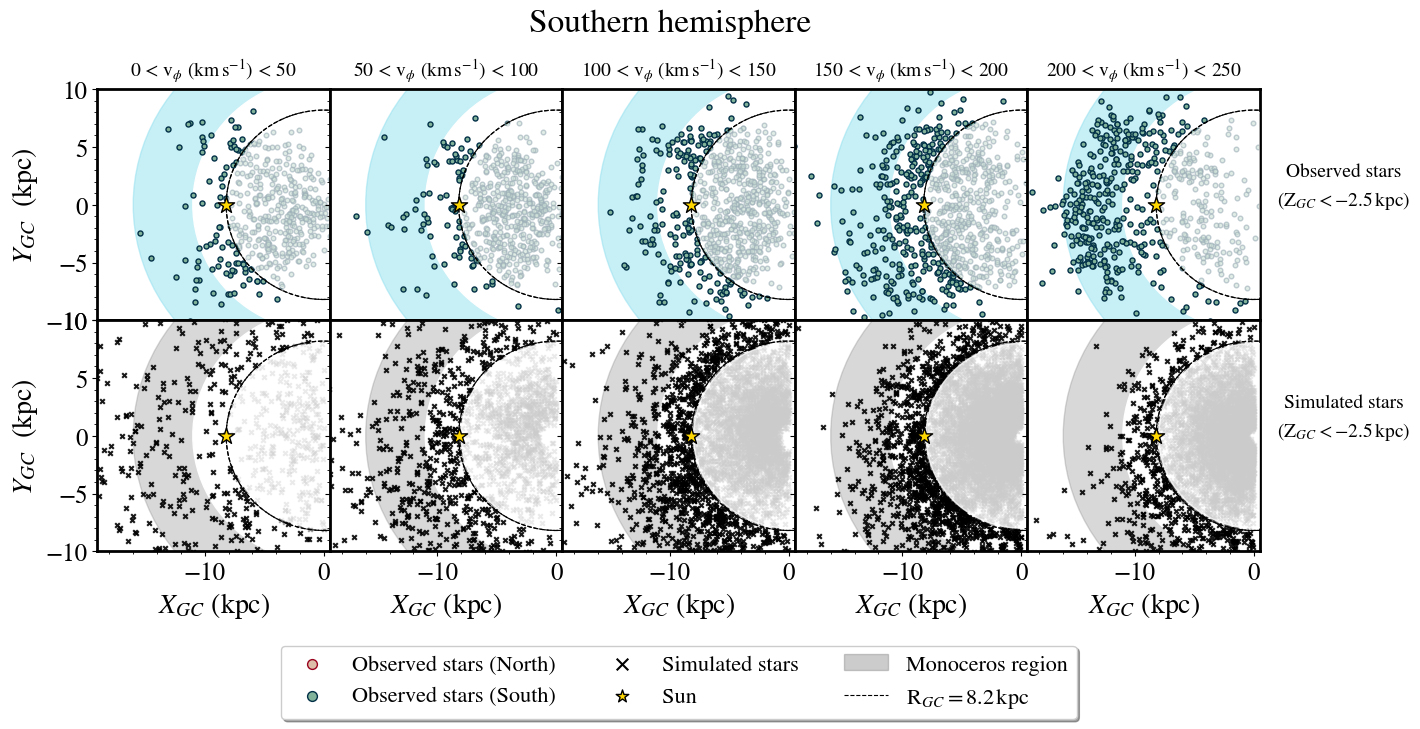}
        \caption{Spatial distribution in the $X_{GC}$--$Y_{GC}$ plane of our M-giant stars, and stars generated by \texttt{Galaxia} after cutting in height, separated into five intervals: of $v_{\phi}\,{\rm (km\,s^{-1})}$: [0, 50], [50, 100], [100, 150], [150, 200], and [200, 250]. The two upper panels refer to northern hemisphere data with $Z_{GC} > 2.5$\,kpc (\textit{M-giant sample} and simulated data, respectively), while the two lower panels show southern hemisphere data with $Z_{GC} < -2.5$\,kpc (\textit{M-giant sample} and simulated data, respectively). The shaded band indicates the Galactocentric radius band where we can find Monoceros, being $12 < R_{GC} {\rm(kpc)} < 17$ in the north and $11 < R_{GC} {\rm(kpc)} < 16$ in the south \citep{newberg2002, Rocha_Pinto2003, morganson2016}. The dashed line indicates the Sun's Galactocentric radius $R_{GC} = 8.2$\,kpc. The yellow star indicates the position of the Sun ([$X_{GC}$, $Y_{GC}$] = [$-$8.2, 0.0]\,kpc).}
        \label{XY_vphi}
    \end{figure*}

\begin{figure*}[ht!]
        \centering
        \includegraphics[width=2.1\columnwidth]{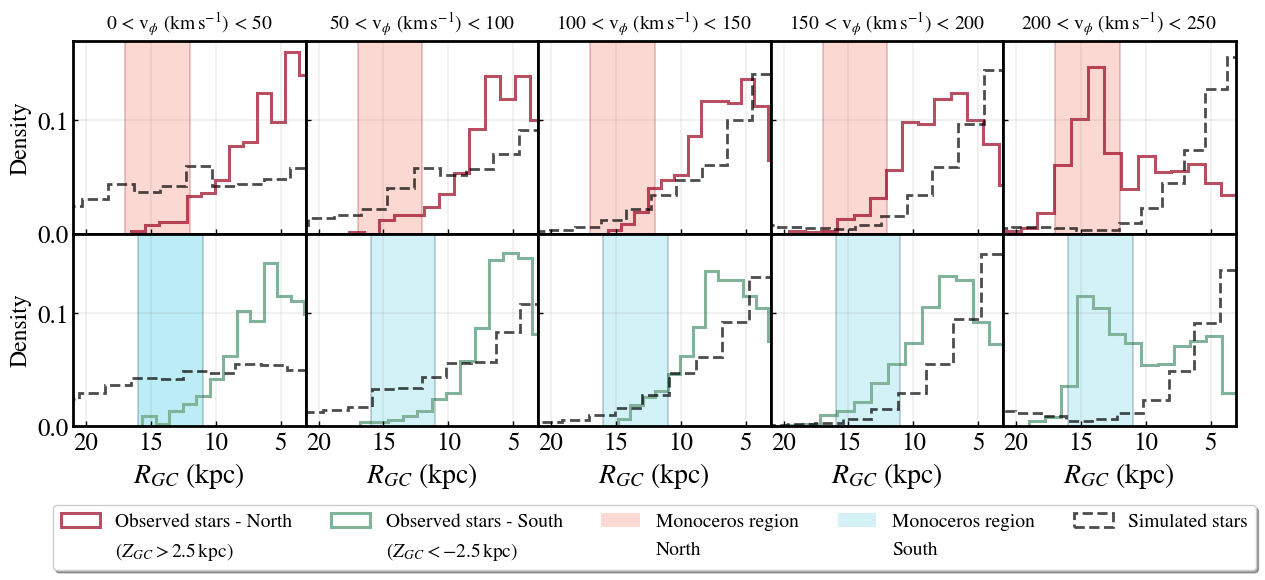}
        \caption{Histograms of Galactocentric distances for M-giant stars shown in Figure \ref{XY_vphi}. The upper panels show data from the northern hemisphere with $Z_{GC} > 2.5$\,kpc, while the lower panels show data from the southern hemisphere with $Z_{GC} < -2.5$\,kpc. The dashed black line shows the distribution of the simulated stars by \texttt{Galaxia}. The red and blue shaded areas represent the Galactocentric radius range where the Mon-N and Mon-S, respectively, are expected to be located.}
        \label{hist_vphis}
\end{figure*}

\subsection{Kinematic signature of Monoceros' stars}
\label{sec:Mon_kinematics}

\vspace{0.1cm}

In order to distinguish the stars of Mon-S and Mon-N structures from other stars of the outer disk, we first analyzed the kinematics of the stars in our \textit{North} and \textit{South M-giant samples}. To facilitate comprehension, we compared the observed data with simulated data generated by \texttt{Galaxia}. This approach aimed to evidence the differences between the expected and observed stellar content.

In Figure \ref{XY_vphi}, we observe a decreasing trend in the number of stars with increasing Galactocentric radius for the first four panels in the observed data in the Northern and Southern hemisphere. The low number of stars beyond $R_{GC} \gtrsim 10$\,kpc is partly due to the completeness of our sample and the low number of objects at this distance. In contrast, the fifth panel displays an unexpected excess of stars that coincides with the Monoceros region (shaded area) in both hemispheres, despite the sample being more than 2.5 kpc away from the Galactic plane
Besides that, when compared to the distribution of simulated stars from the mock catalog (panels with black marks), we note that the stars of both hemispheres do not follow the expected Galactic trend, with the largest fraction of stars being located in the shaded (Monoceros-like) area. At even higher rotational velocities ($v_\phi > 250$\kms), there are effectively no stars left at this distance.

Figure \ref{hist_vphis} shows histograms of the number of stars in bins of $R_{GC}$ for the same ranges of $v_{\phi}$ as presented in Figure \ref{XY_vphi}. In the first four panels, the histograms confirm the decreasing trend in the number of stars with increasing radius discussed above.  In general, the simulated content (black dashed histograms) follows the distribution of the observed stars, with a slight difference in the range where $v_{\phi}\, < 50 {\rm (km\,s^{-1})}$, which exhibits a flat trend. The histogram of the fifth panel ($200 < v_{\phi}\,{\rm (km\,s^{-1})} < 250$) shows that the number of stars sharply increases with radius, forming a peak at $R_{GC} = 13{-}14$\,kpc, revealing the presence of the Monoceros structure in both hemispheres. At these specific values of $Z_{GC}$ and within the velocity range of $200 < v_{\phi}\,{\rm (km\,s^{-1})} < 250$, we can select the Mon-S and Mon-N structures with low contamination. We note that the peaks of distributions within this distance and velocity range is more prominent in Mon-S and Mon-N compared to the inner regions of the Galaxy. 
We want to highlight that the predicted content for a Galaxy with warp and flare (black dashed line), as described by \texttt{Galaxia} code, does not exhibit the structures of Mon-S and Mon-N, despite the larger number of objects in the mock catalog.

\begin{figure}
        \centering
        \includegraphics[width=\columnwidth]{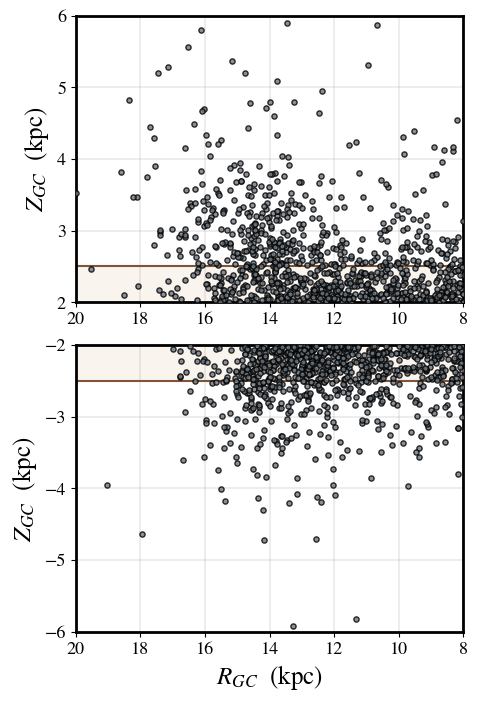}
        \caption{Spatial distribution in the $R_{GC}$--$Z_{GC}$ plane of M-giant stars within $200 < v_{\phi}\,{\rm (km\,s^{-1})} < 250$ and with $R_{GC} > 8$\,kpc. The upper and lower panels contains the stars from the northern and southern hemisphere, respectively. The shaded area indicates the stars with $|Z_{GC}| < 2.5$\,kpc.}
        \label{hemisferios}
\end{figure}

The observed features in Figures \ref{XY_vphi} and \ref{hist_vphis} can be further examined by analyzing their radial and vertical distribution. Figure \ref{hemisferios} shows the projection of stars with a velocity range between $200 < v_{\phi}\,{\rm (km\,s^{-1})} < 250$ in the $R_{GC}$--$Z_{GC}$ plane.  
We note the clear asymmetry between both hemispheres for this sample. Additionally, it can be observed that as the Galactocentric radius increases, the stars exhibit greater heights relative to the Galactic plane, accentuated in the northern hemisphere. However, it becomes challenging to identify clear density patterns at lower $|Z_{GC}|$ values, especially when approaching the Galactic plane, where the characteristic signal of the overdensity becomes less pronounced.

\subsection{Mon-N and Mon-S Distance}
\label{sec:Mon_dist}

\begin{figure*}[ht!]
        \centering
        \includegraphics[width=2\columnwidth]{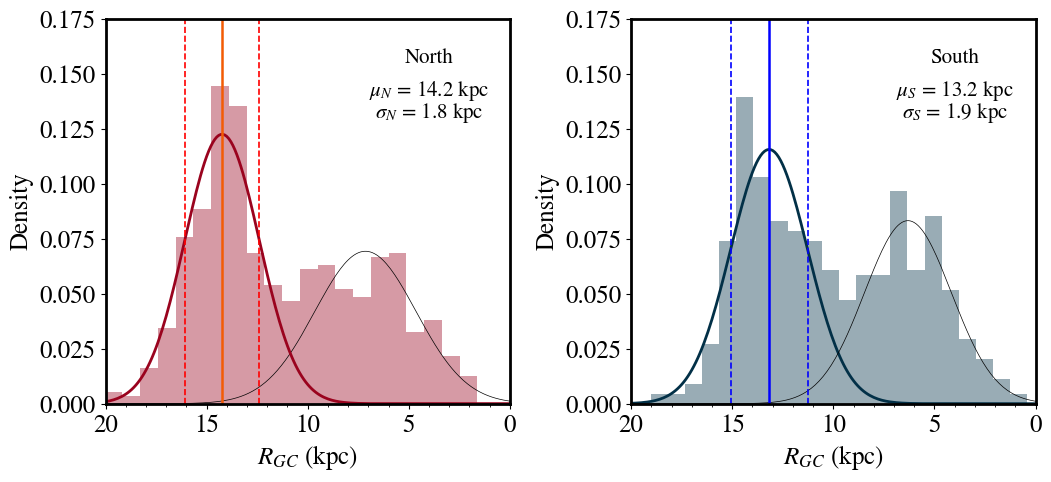}
        \caption{Histogram of star density by Galactocentric radius ($R_{GC}$) for the M-giant sample within $200 < v_{\phi}\,{\rm (km\,s^{-1})} < 250$. The left panel shows the northern hemisphere, and the right panel shows the southern hemisphere. Gaussians were fitted using the GMM method. The vertical solid line indicates the mean ($\mu$) and the vertical dashed line indicates the standard deviation ($\sigma$) of the Gaussian related to the peak associated with Monoceros.}
        \label{GMM_Mon}
  \end{figure*}

The spatial projection of stars using their $v_{\phi}$ values has offered a straightforward approach for identifying the Mon-S and Mon-N structures at high $|Z_{GC}|$ values. Through this selection, it becomes feasible to estimate the distances to both structures.

We employed Gaussian Mixture Models (GMM) to estimate the peak and dispersion of the distance distribution. GMM is a combination of N Gaussian distributions to represent a probability distribution. To perform this analysis, we utilized the \texttt{GaussianMixture} package from \texttt{scikit-learn} \citep{pedregosa2011}, employing the expectation-maximization algorithm \citep{dempster1977} to search for the best-fit model. We found that the optimal number of Gaussian components for both regions is two. For this analysis, we considered only stars within $200 < v_{\phi}\,{\rm (km\,s^{-1})} < 250$ in both hemispheres, as Monoceros is prominently observed within this range. 

Figure \ref{GMM_Mon} clearly illustrates that a two-component Gaussian mixture describes the stellar distribution in our sample. We find that Mon-N (left panel) has a peak distance of $14.2$\,kpc with a wide distribution between approximately $12.4$\,kpc and $16.0$\,kpc. In the case of Mon-S (right panel), the maximum estimated distance is $13.2$\,kpc and a standard deviation of $1.9$\,kpc, indicating a structure ranging from 11.3 to 15.1\,kpc.

In order to  understand the impact of measurement errors on the estimation of Monoceros' location, we employed a Monte Carlo algorithm to generate 1000 realizations of each star's distance, taking into account associated uncertainties. We, then, perform the GMM distance estimation method to each iteration and verify that the mean distance of Monoceros' stars obtained for the Gaussian fits are always within $\leq$5\% of the nominal value. Therefore, we found a final location for the Mon-N peak of 14.24$\pm$0.04\,kpc. For Mon-S, our obtained central location is 13.18$\pm$0.05\,kpc.

We observe that our estimate of the Mon-S peak appears to be offset by approximately 1\,kpc from the peak in stellar density shown in Figure \ref{GMM_Mon} due to the use of a Gaussian to fit the density distribution. However, the distance range we find for Mon-S is consistent with previous studies (e.g., \citealt{newberg2002, xu2015, morganson2016}). Regarding Mon-N, we note that the distance we have found may not represent the actual peak of the overdensity. It is consistent with the beginning of the structure in the northern hemisphere, as presented by \citet{xu2015} and \citet{morganson2016}. This result can be attributed to the sharp drop in the number of M giant stars at $R_{GC} \gtrsim 14.5$ kpc. This limitation arises from the sample selection that remove stars with unreliable phase-space information, primarily due to distance uncertainties and the absence of radial velocity (see Section \ref{sec:data}). As a result, we could not accurately evaluate the extent of Mon-S to longer distances due to the incompleteness of our sample. Nonetheless, the selection of members from both structures remains reliable, as the overdensity can still be observed by considering the selection in height from the Galactic plane ($|Z_{GC}|$) and $v_{\phi}$. 

\begin{figure*}[ht!]
        \centering
        \includegraphics[width=2\columnwidth]{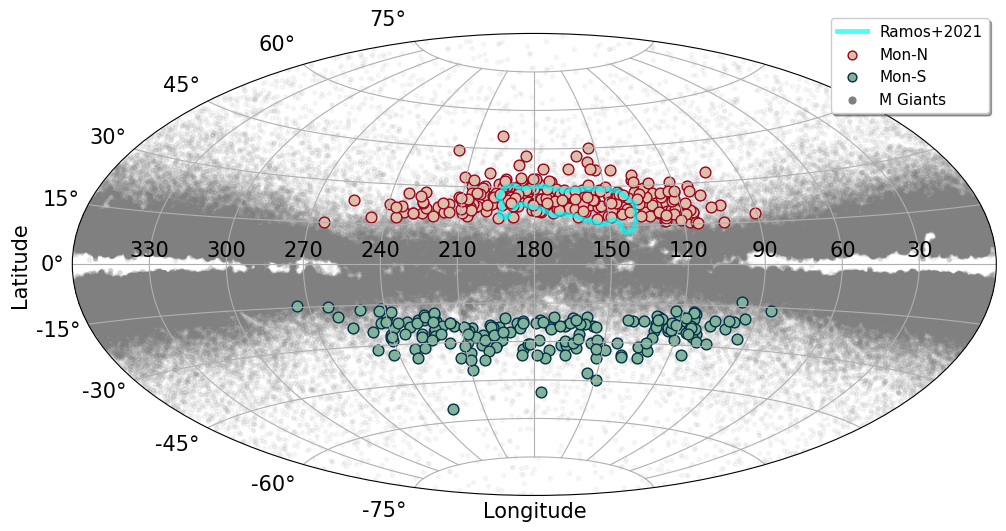}
        \caption{ \textit{Aitoff} projection in Galactic coordinates ($l$ and $b$) of the giant M stars selected as members of the structure of Monoceros. The cyan contour represents the Monoceros structure identified by \citet{Ramos2021} and 
        the background shows the projection of all M-giant stars in our catalog.}
        \label{proj_aitoff}
    \end{figure*}

\subsection{Final selection}
\label{sec:Mon_final_selection}

The analyses presented in Sections \ref{sec:Mon_kinematics} and \ref{sec:Mon_dist}, in combination with the $Z_{GC}$ cuts, allowed us to find member stars of Monoceros with good reliability and, presumably, low contamination. The final designed criteria for selecting Mon-S and Mon-N members can be summarized as follows:

\begin{itemize}
    \item Mon-N: $Z_{GC} > 2.5$\,kpc, $200 < v_{\phi}\,{\rm (km\,s^{-1})} < 250$ and $12.4 < R_{GC}\text{(kpc)} < 16.0$;

    \item Mon-S: $Z_{GC} < -2.5$\,kpc, $200 < v_{\phi}\,{\rm (km\,s^{-1})} < 250$ and $11.3 < R_{GC}\text{(kpc)} < 15.1$.
\end{itemize}

Our selection criteria resulted in the identification of 264 M-giant star candidates belonging to Mon-N, which are situated within the Galactic coordinates range of 91$^{\circ}$ $< l <$ 264$^{\circ}$ and 15$^{\circ}$ $< b <$ 50$^{\circ}$. These stars have a median height of $| Z_{GC} |$ = 3.0\,kpc and a maximum of $| Z_{GC} |$ = 7.5\,kpc in relation to the Galactic plane.  In relation to Mon-S, we identified 182 M-giant stars with Galactic coordinates ranging from 85$^{\circ}$ $< l <$ 275$^{\circ}$ and $-55^{\circ} < b < -14^{\circ}$, reaching a median height of $| Z_{GC} |$ = 2.8\,kpc and a maximum of $| Z_{GC} |$ = 5.9\,kpc.  In both Mon-S and Mon-N regions, we have not identified any contamination from halo stars according to the analysis of their energy and angular momentum. Additionally, when examining the Toomre diagram, we find that less than 1$\%$ of stars exhibit velocities ($\sqrt{v_r^2 + v_z^2}$) greater than 100 ${\rm km\,s^{-1}}$, which is consistent with the characteristics typically associated with nearby thick disk stars \citep[e.g.,][]{Venn2004, Bensby2014}. It is worth noting that the $v_{\phi}$ range observed in Mon-N and Mon-S stars is typically associated with thin disk stars. However, it is unusual to find such a significant number of thin disk stars at this distance and height from the Galactic plane, as can be seen in Figures 3 and 4.

Figure \ref{proj_aitoff} shows the aitoff projection of the M-giant stars selected as members of the Monoceros overdensity. The Mon-N stars from our selection are overlaid with the selections made in previous studies by \citet{Ramos2021}. It is important to mention that the distributions of Mon-S and Mon-N may extend to other values of longitude \citep[e.g.,][]{morganson2016}. However, due to the incompleteness of our sample at the distance of Monoceros in directions other than $110^\circ > l > 250^\circ$, we are unable to assess the complete distribution of the structure.


\begin{figure*}[ht!]
        \centering
        \includegraphics[width=2.1\columnwidth]{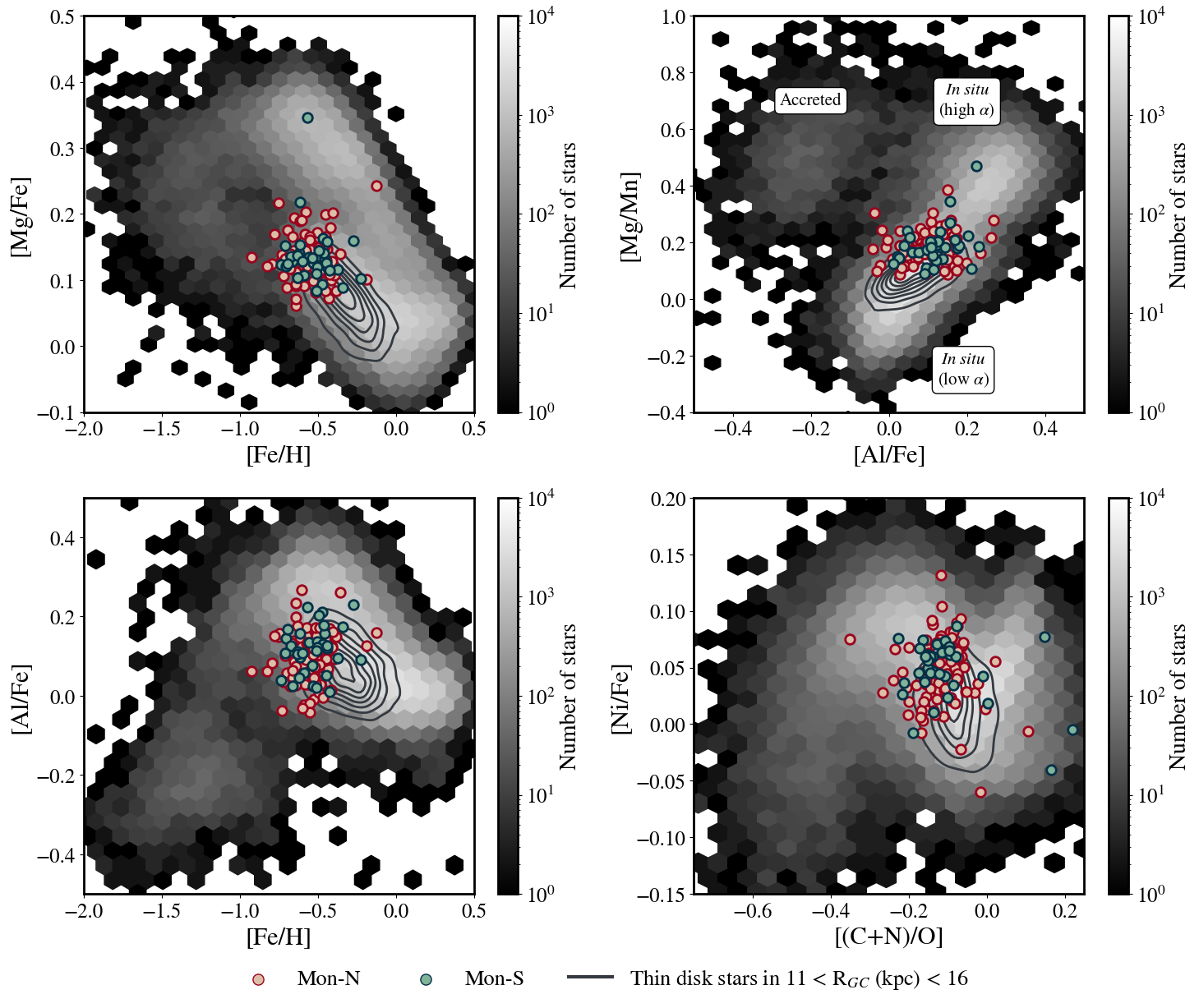}
        \caption{Chemical abundances of Mon-N, Mon-S stars in the [Mg/Fe]--[Fe/H] (top right), [Mg/Mn]--[Al/Fe] (top right), [Al/Fe]--[Fe/H] (bottom-left) and [Ni/Fe]--[(C+N)/O] (bottom right). 
        The background 2D histogram shows the distribution of the APOGEE stars. The contour lines represent the region occupied by thin disk stars located in the intervals $11 < R_{GC} (\text{kpc}) < 16$ and $-1 < Z_{GC} (\text{kpc}) < 1$. }
        \label{quimica}
    \end{figure*}
    
\section{Chemical Abundances}
\label{sec:abundances}

The chemical properties of the Mon-S and Mon-N regions have received relatively little analysis. While \citet{laporte2020} presented and analyzed the [Mg/Fe]--[Fe/H] space for Mon-N, Mon-S remains individually unexplored. Therefore, in this section, we will present the analysis of the chemical abundance patterns of Mon-N and Mon-S in various abundance spaces, and discuss their origin scenarios.

The \textit{M-giant sample} lack elemental abundances that would enable us to study the chemical properties of the overdensities. Therefore, we applied the criteria outlined in Section \ref{sec:Mon_final_selection} to select Mon-S and Mon-N stars from the \textit{APOGEE sample} (see Section \ref{sub:apogeesample}). Our selection yielded 99 stars in Mon-N and 34 stars in Mon-S.

Figure \ref{quimica} shows the chemical abundances of stars belonging to the Mon-N and Mon-S structures in various abundance spaces. 
The top left panel shows the [Mg/Fe]--[Fe/H] space, in which it is possible to see the separation between the halo, thick disk, and thin disk \citep{Hawkins2015,mack2019acc,das2020}. The stars from Mon-N and Mon-S clearly overlap the region expected for the outer thin-disk stars. The top right panel presents the [Mg/Mn]--[Al/Fe] space and shows that the Monoceros stars overlap with the region expected for \textit{in situ} stars with low $\alpha$ elements. The bottom panels show that our samples overlap with the disk region (denser region) also in [Al/Fe]--[Fe/H] and [Ni/Fe]--[(C+N)/O]. The results indicate that both Mon-S and Mon-N exhibit chemical compositions similar to that of the outer disk. This finding is consistent with the results obtained by \citet{laporte2020} for the northern portion of Monoceros and for other more distant stellar overdensities near to the disk \citep{bergemann2018, hayes2018triand, Abuchaim2023}.

It is noteworthy that there is an overlap between Mon-S and Mon-N in all panels. The average [Fe/H] for Mon-N is approximately $-$0.55, while for Mon-S, it is around $-$0.53. 
These results further support the notion that both groups are part of the same, larger substructure of \textit{in situ} origin.

To compare Mon-S and Mon-N stars from the \textit{APOGEE sample} with characteristic outer thin disk stars, we select stars from the \textit{APOGEE sample} falling within the same distance range as Mon-S and Mon-N (as detailed in Section \ref{sec:Mon_final_selection}). Furthermore, we restricted ourselves selecting stars located within the distance range of $-1.0 < Z_{GC} ({\rm kpc}) < 1.0$. In all panels of Figure \ref{quimica}, the black lines correspond to isodensity contours representing the APOGEE 
outer thin disk stars. We note that the stars in Mon-S and Mon-N exhibit a narrower range of chemical composition compared to the outer thin disk stars. In this way, the presence of these stars with this chemical profile at significant distances from the Galactic plane suggests that the stars in Mon-S and Mon-N underwent a heating process before the prolonged formation of the outer disk.


\section{SUMMARY}
\label{sec:conclusion}

In this study, we employ a comprehensive approach combining data from 2MASS, AllWISE, Gaia, and APOGEE, along with StarHorse code distances, to chemically and kinematically characterize the Monoceros overdensity in the southern and northern hemispheres and investigate their origin and potential connection. To reinforce our interpretations, we generated a mock catalog with the \texttt{Galaxia}, enabling us to contrast the predicted content with observed data.

To select Monoceros stars, we utilized an M-giant star catalog derived from the 2MASS, AllWISE, and Gaia surveys. By applying a simple kinematic criterion, for stars located at $|Z_{GC}| = 2.5$\,kpc, we were able to identify Mon-S and Mon-N, within a well-defined velocity range (200 $<$ $v_{\phi}\,{\rm (km\,s^{-1})} <$ 250). This velocity range is consistent with the typical velocities of thin disk stars, which is surprising considering that they have been heated to these high $|Z_{GC}|$ values. By implementing this selection criterion, we determined the Galactocentric distances of Mon-N and Mon-S as $R_{GC}$ = $14.2 \pm 1.8$\,kpc and $R_{GC}$ = $13.2 \pm 1.9$\,kpc, respectively. It is worth noting that the determination of distances for Mon-S and Mon-N is biased by the incompleteness of our M giant sample at large distances. However, this limitation does not affect the kinematic and chemical analysis of the members of Monoceros. Our method of selection of Monoceros structures can be applied in the future in a more complete sample, aiming at a new estimate of the distance of Mon-N and Mon-S.

Additionally, it is evident that these structures are not symmetrical and do not extend to the same distance from the Galactic plane (see Figure \ref{hemisferios}). Therefore, it is more challenging to identify a clear continuation of Mon-N and Mon-S structures at lower $Z_{GC}$ values. We also note that the structures that we studied have very similar Galactocentric radius and azimuthal velocity to the anticentre ridges discovered in \citet{Antoja2021}. However, the Mon structures are at significantly higher $Z_{GC}$. For the same reasons discussed above, we could not trace a possible continuity of our structures with these ridges.

Using the selection criteria for Mon-S and Mon-N, we successfully were able to identify stars from these stellar overdensities in the APOGEE spectroscopic survey with available chemical information. Through an analysis of various parameter spaces, including [Mg/Fe]--[Fe/H], [Mg/Mn]--[Al/Fe], [Al/Fe]--[Fe/H], and [Ni/Fe]--[(C+N)/O], we characterized the detailed abundance patterns of Mon-S for the first time and confirmed the similarity in chemical abundances between Mon-S and Mon-N, consistent with stars typically found in the thin disk of our Galaxy. It is noteworthy that stars exhibit a narrower range of chemical abundances compared to the outer thin disk stars. It implies that these stars were dynamically heated through an event prior to the chemical enrichment. This finding suggests that the heating of Mon-S and Mon-N stars is more likely a result of tidal interactions with other satellite galaxies, rather than an extended secular evolution process. The small range in chemical compostion of Monoceros stars could be linked to perturbations caused by the passage of the Sgr dSph. 

In summary, this study provides robust constraints on the spatial distribution, kinematics, and chemical abundances of Mon-S and Mon-N. The combination of kinematic and chemical abundance analyses supports the hypothesis that Mon-S and Mon-N were formed \textit{in situ} and likely by the same mechanism. These findings are in concordance with simulations that consider the influence of Sagittarius (Sgr) dwarf spheroidal (dSph) galaxy and predict the existence of stellar overdensities above and below the Galactic disk at various distances \citep[e.g.,][]{gomez2016, laporte2018}. However, due to our sample consisting of giant stars, we do not have reliable age information to address on which pericenter passage of the Sgr dSph these structures could have originated \citep[see][]{das2023}. In the future, with the next generation of surveys, we hope be able to better estimate the stellar ages and obtain more information on the epoch of the formation of the outer disk overdensities.

\acknowledgements

All authors are indebted to those involved with the ``Brazilian Milky Way group meeting" for weekly constructive discussions. L.B. also thanks Fabr\'icia O. Barbosa for her contribution to the calculation of the error estimates and Jo\~ao A. Amarante for conversations that contributed to improve the quality of this work. 

L.B. acknowledges CAPES/PROEX (proc. 88887.821814/2023-00) and FAPESP (proc. 2021/09967-8). H.D.P. also thanks FAPESP (procs. 2018/21250-9 and 2022/04079-0). S.R. thanks partial financial support from FAPESP (procs. 2015/50374-0 and 2020/15245-2), CAPES, and CNPq. G.L. acknowledges FAPESP (procs. 2021/10429-0 and 2022/07301-5).  A.P.-V. acknowledges the DGAPA–PAPIIT grant IA103122. TA acknowledges the grant RYC2018-025968-I funded by MCIN/AEI/10.13039/501100011033 and by ``ESF Investing in your future''.  This work was (partially) supported by the Spanish MICIN/AEI/10.13039/501100011033 and by "ERDF A way of making Europe" by the “European Union” and the European Union «Next Generation EU»/PRTR,  through grants PID2021-125451NA-I00 and CNS2022-135232, and the Institute of Cosmos Sciences University of Barcelona (ICCUB, Unidad de Excelencia ’Mar\'{\i}a de Maeztu’) through grant CEX2019-000918-M. R.M.S ackonowledges CNPq (proc. 306667/2020-7).

This work has made use of data from the European Space Agency (ESA) mission {\it Gaia} (\url{https://www.cosmos.esa.int/gaia}), processed by the {\it Gaia} Data Processing and Analysis Consortium (DPAC, \url{https://www.cosmos.esa.int/web/gaia/dpac/consortium}). Funding for the DPAC has been provided by national institutions, in particular the institutions participating in the {\it Gaia} Multilateral Agreement.

Funding for the Sloan Digital Sky Survey IV has been provided by the Alfred P. Sloan Foundation, the U.S. Department of Energy Office of Science, and the Participating Institutions. SDSS-IV acknowledges support and resources from the Center for High Performance Computing  at the University of Utah. The SDSS website is \url{www.sdss.org}. SDSS-IV is managed by the Astrophysical Research Consortium for the Participating Institutions of the SDSS Collaboration. 

\clearpage

\bibliographystyle{aasjournal}

\bibliography{bibliography.bib}




\end{document}